\documentclass[11pt,prl,floatfix,superscriptaddress,amsmath,amssymb,aps]{
revtex4}
\usepackage{epsfig}
\usepackage{bm}
\usepackage{amsmath,wasysym} 
\usepackage[greek,english]{babel}
\usepackage{teubner}
\usepackage[normalem]{ulem}

\RequirePackage{color}
\definecolor{MyDarkGreen}{rgb}{0.02,0.60,0.06}

\newcommand{\blue}[1]{{\color{blue}{#1}}}

\def\be{\begin{equation}}
\def\ee{\end{equation}}
\def\bey{\begin{eqnarray}}
\def\eey{\end{eqnarray}}
\def\vac#1{{\bf{#1}}}
\def\bmu{{\pmb{\mu}}}
\def\bnabla{{\pmb{\nabla}}}
\def\qq{{\hbox{\foreignlanguage{greek}{\coppa}}}}
\def\qqq{{\hbox{\foreignlanguage{greek}{\footnotesize\coppa}}}}

\begin{document}

\title{On the role of Fourier modes in  finite-size scaling above the upper critical dimension}

\author{Emilio J. Flores-Sola} 
\affiliation{Applied Mathematics Research Centre, Coventry
University, Coventry CV1 5FB, United Kingdom}
\affiliation{Institut Jean Lamour, CNRS/UMR 7198, Groupe de Physique
Statistique, Universit\'e de Lorraine, BP 70239, F-54506 Vand\oe uvre-les-Nancy Cedex, France}
\affiliation{{Doctoral College for the Statistical Physics of Complex Systems,  Leipzig-Lorraine-Lviv-Coventry $({\mathbb L}^4)$}}
\author{Bertrand Berche}
\affiliation{Institut Jean Lamour, CNRS/UMR 7198, Groupe de Physique
Statistique, Universit\'e de Lorraine, BP 70239, F-54506 Vand\oe
uvre-les-Nancy Cedex, France}
\affiliation{{Doctoral College for the Statistical Physics of Complex Systems,  Leipzig-Lorraine-Lviv-Coventry $({\mathbb L}^4)$}}
 \author{Ralph Kenna}
\affiliation{Applied Mathematics Research Centre, Coventry
University, Coventry CV1 5FB, United Kingdom}
\affiliation{{Doctoral College for the Statistical Physics of Complex Systems,  Leipzig-Lorraine-Lviv-Coventry $({\mathbb L}^4)$}}
 \author{Martin Weigel}
\affiliation{Applied Mathematics Research Centre, Coventry
University, Coventry CV1 5FB, United Kingdom}
\affiliation{{Doctoral College for the Statistical Physics of Complex Systems,  Leipzig-Lorraine-Lviv-Coventry $({\mathbb L}^4)$}}
\date{\today}

\begin{abstract}
Renormalization-group theory stands, since over 40 years, as one of the pillars of  modern physics. 
As such, there should be no remaining doubt regarding its validity.
However, finite-size scaling, which derives from it, has long been poorly understood  above the upper critical dimension {$d_c$} in models with free boundary conditions.
Besides its fundamental significance for scaling theories, the issue is  important at a practical level because 
finite-size, statistical-physics systems, with 
free boundaries  {above $d_c$,} are experimentally accessible {with long-range interactions.}
Here we address the roles played by Fourier modes for such systems and show that the current phenomenological picture is not supported for all thermodynamic observables
either with  free or periodic boundaries.
Instead, the correct picture emerges from   a sector of the renormalization group hitherto considered unphysical.
\end{abstract}


 \maketitle


In this Letter we  address a subtle question, still open in the theory of finite-size scaling (FSS). 
Although it may cursorily appear  an academic exercise, this question has to be decisively and convincingly resolved {since} it concerns the very foundations of the renormalization group (RG), one of the greatest achievements of theoretical physics~\cite{Wi71}. 
Given the nature of our investigation, and with the aim of {a} comprehensive presentation, 
we  {emphasize essential}  historical and relatively technical  elements {of the theory} before introducing the problem {of} Fourier modes in the treatment of  boundary conditions. 

FSS {is a well} developed component of  modern theories of critical phenomena.
It was initially introduced on heuristic grounds~\cite{Fi71} and then  understood within the frame of Wilson's {RG}~\cite{Wi71,WiKo74,Ma76,Br82}. 
{The system's} inverse typical linear scale $L^{-1}$ appears like the reduced temperature $t=(T-T_c)/T_c$ or  magnetic field $h=H/T_c$ as a scaling field governing {flow} towards the RG fixed points (FP) which control the critical singularities {of t}hermodynamic properties,
\bey
&&C(t)\sim|t|^{-\alpha},\qquad m(t) {\rm{\raisebox{-1.0ex}{ {\small 
\shortstack{$\sim$ \\ {\tiny{$t\!\!<\!\!0$}}}} } }} 
|t|^\beta,\qquad m(h)\sim 
|h|^{1/\delta},\label{eq-cmm} \\
&&\chi(t)\sim |t|^{-\gamma},\qquad \
\xi(t)\sim|t|^{-\nu},\qquad g(\vac x)\sim |\vac 
x|^{-(d-2+\eta)}.\label{eq-chixig}
\eey
Here, $C$, $m$, $\chi$, $\xi$ and $g$ are, respectively, the singular parts of the specific heat, magnetization, susceptibility, correlation length and correlation function, and the fields $t$ or $h$ when not specified are zero.
The long-distance properties at the {FP's} {depend} on space dimension {$d$} and order-parameter symmetry, {but not on} details such as the {short range of} microscopic interactions,  or lattice symmetry. Hence, there exists a set of properties (the critical exponents 
as well as combinations of critical amplitudes, not considered here) which  {\em rigorously} take the same values for  different systems. Magnetic systems, for example, can share these properties with fluids. This is the meaning of the term ``universality'' used in this context~\cite{Gr70}.

The usual phenomenological argument for FSS (which turns out to be valid only below the upper critical dimension)  can be summarized as follows~\cite{Fi71,Br82}. 
Assume { a  quantity $P$ exhibits a  singularity 
in the vicinity of the critical point $T_c$, measured} by a critical exponent $\rho$, so that $P(t)\sim|t|^{\rho}$. 
Then, the singularity {develops} in a finite-size system as
\be 
 P(t,L^{-1})\sim P(t,0) \left[{ {L} / {\xi(t)}}\right]^{-\frac{\rho}{\nu}},
 \label{FSSPictureStandard}
\ee
where, it is argued, {the ratio} $L/\xi(t)$ appears because {it involves} the only two length scales governing long-distance behavior, and  the power {on the right-hand side}  is such that the singularity in $t$ is washed out at $T_c$ for the finite system. 
This argument predicts a FSS exponent $\gamma/\nu$ for the susceptibility, and $-\beta/\nu$ for the magnetization.

A spectacular result of Wilson's RG is the explanation for the existence of an upper critical dimension $d_c$ above which Landau mean-field theory (MFT) is recovered, with 
\bey&
\alpha& = 0,\qquad \beta=1/2,\qquad  \delta=3,\qquad \label{eqMFT1} 
\\
&
\gamma&=1,\qquad \nu=1/2,\qquad \eta=0\label{eqMFT2} \eey 
for $\phi^4$-field theory.
To fix ideas, let us consider a nearest-neighbour Ising model  consisting of spins $s_{\vac x}$ located on the sites $\vac x$ of a regular hypercubic lattice {with} unit vectors $\bmu$ ($|\bmu|=a$, the lattice spacing). The exact partition function
\be
Z=\sum_{ \{ s_{\vac x} \} } \exp{ \left[
\beta\left(
J\sum_{\vac x}\sum_{\bmu}s_{\vac x}s_{\vac x+\bmu}+H\sum_{\vac x}s_{\vac x}
\right) \right]
}
\label{eq-Z}
\ee
can be rewritten via  a Hubbard-Stratonovich transformation as a functional integral {over} $\phi(\vac x)\in\mathbb R$, $Z \simeq  \prod_{\vac x}\int d\phi(\vac x)\ \!
 e^{-S[\phi]}$, {with}
\bey
 S[\phi]& =& {\int d^dx
 \left({
f_0+
\frac{r_0}{2}\phi^2
+\frac{u}{4}\phi^4
+\frac{c}{2}|\bnabla\phi
|^2-h\phi
}\right)},\ \label{eq-action}
\eey
{where} $f_0=-a^{-d}\ln 2$, $r_0=a^{-2}t$, $u=2a^{d-4}$, $c=(2d)^{-1}$ and 
$h=\beta a^{-(d/2+1)} H$. Dimensional analysis shows that the free energy $f_0$ has scaling dimension $d$, the coefficient $r_0$, proportional to the reduced temperature, has dimension 
$y_t=2$, the coefficient of the quartic term $u$ has  $y_u=4-d$ and the magnetic field 
has 
$y_h=d/2+1$.
The eigenvalues $y_i>0$ control the  flow of the relevant fields at the {FP}, leading to {homogeneity:}
\bey
f(t,h)&=&b^{-d}F(b^{y_t}t,b^{y_h}h),\label{eq-fG}\\
\xi(t,h)
&=& b\Xi(
b^{y_t}t,b^{y_h}h
),
\label{eq-xiG}\\
g(t,h,\vac x)
&=& b^{-d+2}G(
b^{y_t}t,b^{y_h}h,b^{-1}\vac x).
\label{eq-gG}
\eey 
The critical exponents in Eqs.~(\ref{eq-cmm})-(\ref{eq-chixig}) emerge through {the} scaling laws, 
\bey&
\alpha& \!\!\! = (2y_t-d)/y_t,\qquad \! \beta=(d-y_h)/y_t,\qquad  \! \! 
\delta=y_h/(d-y_h),~\quad  \label{eqRG1} 
\\
&
\gamma& \!\! \!=(2y_h-d)/y_t,\qquad \! \! \nu=1/y_t,\qquad \ \ \qquad
\eta=d-2y_h+2.\label{eqRG2} \eey

Above $d_c=4$, $y_u<0$,  {and} one expects that  critical behavior  should be controlled by the Gaussian FP   $(t,h,u)=(0,0,0)$.
But  a discrepancy between the MFT exponents in Eqs.~(\ref{eqMFT1}) and Eqs.~(\ref{eqRG1}) indicates that the limit $u\to 0$ has to be taken with care; $u$ is a {\em dangerous irrelevant variable} (DIV)~{\cite{DIV}}, at least, as was  first thought, in the free-energy sector. The agreement between Eqs.~(\ref{eqMFT2}) and (\ref{eqRG2}) suggests that there is no {danger}  in the sector of the correlations~\cite{Cardy}.
A proper inclusion of this mechanism   resolves the discrepancy in the free-energy sector.
In Fourier space, for a periodic system, the quadratic part of the action (\ref{eq-action}) reads  $\frac12\sum_{\vac k}(|\vac k|^2+r_0)|\phi_{\vac k}|^2$, and the quartic term $\frac14u\int d^dx\ \!\phi^4(\vac x)=\frac{u}{4L^d}\sum_{\vac k_1\vac k_2\vac k_3}\phi_{\vac k_1}\phi_{\vac k_2}\phi_{\vac k_3}\phi_{-\vac k_1-\vac k_2-\vac k_3}$
can be expanded as $\frac 14\frac {u}{L^d}\phi_0^4+\frac32\frac{u}{L^d}\phi_0^2\sum_{\vac k}|\phi_{\vac k}|^2$ up to higher-order corrections  in the non-zero modes, such that the action can be approximated by
\be
S[\phi]\simeq{ \frac 12\Bigl({r_0}+ \frac{3u}{2L^d}\sum_{\vac k\not=0} 
|\phi_{\vac k}|^2\Bigr)
 \phi_0^2
 +\frac{u}{4L^d}\phi_0^4}{+\frac12\sum_{\vac k\not=0}(r_0+c|\vac 
k|^2)|\phi_{\vac k}|^2
 - {hL^{d/2}\phi_0}.
 }\label{eq-actionexpanded}
\ee
Only the zero mode couples to $h$ and its quartic self-interaction means that $u$ is dangerous.
The non-zero modes  neither couple to $h$ nor have a dangerous quartic term.

The zero mode is thus responsible for  anomalous FSS behavior above $d_c$ and  leads to 
\begin{eqnarray}
    f(t,u,h) = b^{-d}F\left( \frac{b^{y_t}t-b^{y_u}u}{b^{y_u/2}u^{1/2}},\frac{b^{y_h}h}{b^{y_u/4}u^{1/4}}\right) 
.\quad
\label{eq-f}
    \end{eqnarray}
The temperature field $t$ is governed by a modified RG exponent $y_t^*=y_t-\frac 12y_u=\frac d2$ and the magnetic field $h$ by  $y_h^*=y_h-\frac 14y_u=\frac{3d}{4}$~\cite{BZJ85,LuBl97a,BiLuReview}.
{Temperature dependencies} of the magnetization and susceptibility follow by differentiating  (\ref{eq-f}) {wrt} $h$ and choosing the scale factor 
$b=(t/u^{1/2})^{-2/d}$. Hence
\bey
m(t,u,0)&=&B(u)t^{1/2}M\left[{1-A(u)t^{1-4/d}}\right],\label{eq-mt}\\
\chi(t,u,0)&=&\Gamma(u)t^{-1}X\left[{1-A(u)t^{1-4/d}}\right]\label{eq-chit}
\eey
and we now obtain the correct MFT exponents above $d_c$. 
[We omit here the specific heat and critical isotherm for which the same argument {holds;
Eqs.~(\ref{eqRG1})-(\ref{eqRG2})  deliver} {\em all} mean-field exponents with $y_t^*$ and $y_h^*$ in place of the original scaling dimensions.] 
The finite-size behaviour is  immediate by setting $b=L$ in Eqs.(\ref{eq-f}) and differentiating appropriately \cite{BNPY}, e.g. 
\bey
m_{T_c}(u,L^{-1})&\sim& L^{-d/4}M(AL^{2-d/2}u^{1/2}),\label{eqFSSm}\\
\chi_{T_c}(u,L^{-1})&\sim& L^{d/2}X(AL^{2-d/2}u^{1/2}).\label{eqFSSchi}
\eey
If the finite-size correlation length were bounded by the system length {{\cite{BNPY}}}, one could not write the combination {{$L^{y_t^*}t$}}, which enters the free energy, as a ratio  $L/\xi$ along the lines {of} Eq.(\ref{FSSPictureStandard}). 
Within  this framework, another length scale $\ell (t)$ was introduced, {dubbed the thermodynamic length,} with $\ell (t) \sim t^{-2/d}$, FSS being governed by the ratio $L/\ell (t)$ instead \cite{Bi85}.

Contrary  to  previously widespread opinion~\cite{BNPY,Privman,Cardy,Bi85,KBS,Nish,ChaikinLubensky}, 
the correlation sector also needs reexamination and the homogeneity assumption above $d_c$ {there} takes the form~\cite{ourEPL}
\bey
\xi(t,u)
&=& b^{\qqq}\Xi\left(
b^{d/2}\frac{t}{u^{1/2}}-Ab^{2-d/2}u^{1/2} 
\right),
\label{eq-xi}\\
g(t,u,\vac x)
&=& b^{-d/2}G\left(
b^{d/2}\frac{t}{u^{1/2}}-Ab^{2-d/2}u^{1/2}
,b^{-1}{\vac x}
\right),\nonumber\\
\label{eq-g}
\eey
where we omit the {$h$-}dependence for clarity. This leads to a new interpretation, dubbed QFSS in \cite{ourNPB}, above $d=d_c$ dimensions.
[The ``Q'' refers to the introduction of a  pseudocritical exponent $\qq$ which governs the FSS of the correlation length in Eq.(\ref{eq-xi}).]
The finite-size behaviour is transparent
from Eqs.(\ref{eq-xi}) and (\ref{eq-g}); fixing the scale factor $b=L$ we get
\bey
\xi_{T_c}(u,L^{-1})&\sim&L^\qqq\Xi(AL^{2-d/2}u^{1/2}),\label{eqFSSxi}\\
g_{T_c}(u,\vac x,L^{-1})&\sim&L^{-d/2}G(AL^{2-d/2}u^{1/2}).\label{eqFSSg}
\eey
Above the upper critical dimension, $\qq=d/d_c$ and the notion of thermodynamic length is abandoned in the QFSS picture \cite{ourNPB,ourCMP,ourEPL,ourEPJB}.
Below $d_c$, $\qq=1$ and ordinary FSS is recovered.

This picture is  corroborated by {analytical and numerical calculations}   for systems  with periodic boundary conditions (PBC), both in the short-range Ising model (SRIM) above $d_c=4$ and in long-range Ising model  (LRIM) above 
$d_c=2\sigma$~\cite{ourNPB,ourCMP,ourEPL,ourEPJB}.
In the latter case, the general discussion has to be modified. The sum over interactions in Eq.(\ref{eq-Z}) is extended to all pairs with decaying couplings $J_{\vac x-\vac x'}\sim J/|\vac x-\vac x'|^{d+\sigma}$. The MFT critical exponents of Eqs.~(\ref{eqMFT1}) and~(\ref{eqMFT2}) remain valid except that $\nu=1/\sigma$ and $\eta = 2-\sigma$, and the
RG eigenvalues at the Gaussian FP take the form $y_t=\sigma$, $y_h=(d+\sigma)/2$, and $y_u=2\sigma-d$ \cite{FMN72}. {One} recovers the SRIM values with $\sigma=2$. The standard scaling laws (\ref{eqRG2}) are satisfied above $d_c$ while those of (\ref{eqRG1}) are not, indicating again the dangerous irrelevancy of the quartic term in the action, which now contains an additional $|\vac k|^\sigma$ term.
The modified (starred) RG dimensions are the same as before.
The agreement between numerical results {{for PBC's}} and the scaling picture of Eqs.~(\ref{eqFSSm}), (\ref{eqFSSchi}), (\ref{eqFSSxi}) and (\ref{eqFSSg}) is complete when simulations are performed at  $T_c$, but also when they are performed at the pseudocritical point $T_L$ 
(defined by the size-dependent temperature where a quantity such as the susceptibility  exhibits a  maximum).

When FBC's are imposed, there appears {an} intriguing feature. 
Simulation results are  consistent with the above picture at the pseudocritical point $T_L$ of the FBC system \cite{ourNPB}, but not at $T_c$  where, instead, standard FSS with the Landau MFT-type exponents of Eq.(\ref{eqMFT2}) has been obtained {\em for the susceptibility} {{\cite{RuGa85,WiYo14,Watson,LuMa11,LuMa14}. }}

As recently shown by Wittman and Young \cite{WiYo14} (see also \cite{RuGa85,LuBl97a,BiLuReview}), the Fourier modes play a key role.
The non-zero modes, which are not affected by DIV's, contribute to {the} FSS of the susceptibility with the  Landau ratio  $\gamma/\nu=2$  for PBC's  at both $T_c$ and $T_L$
and they argued for the same Landau behaviour of analogous modes for FBC's at $T_L$.
We show below that standard FSS with Landau exponents is not correct; it is in conflict with the RG.

We  follow Ref.\cite{RuGa85} and perform
a sine-expansion of the scalar field in Eq.(\ref{eq-action}) satisfying 
$\phi({\bf x})=0$ at the free surfaces: 
$\phi({\bf x})=\sum_{\bf k}\phi_{\bf k}\prod_{\alpha=1}^d\sqrt{2/L}\sin k_\alpha x_\alpha$, where  
$k_\alpha =n_\alpha \pi/L$, $n_\alpha =1,2,\dots, L$. 
In ${\bf k}-$space, the action takes a form slightly {different to that for}  PBC's 
and one must distinguish  modes for which all $n_\alpha $-values are odd integers.
These are analogous to the zero mode in the PBC case and we denote their set by $Q$. 
We denote the remaining modes by $G$.
The action now reads~\cite{RuGa85}
\begin{eqnarray}
  S[\phi] 
	 = {\frac 12}
	   \sum_{{\bf k}}
              \Bigl({r_0+c|{\bf k}|^2
						 }\Bigr)\phi_{{\bf k}}^2
	  -
		\left({\frac{8}{L}}\right)^{\frac{d}{2}}
		h
		\sum_{\vac k {\in Q}}
		\phi_{{\bf k}}
		\prod_{j=1}^d{\frac{1}{k_j}}
 \nonumber\\
 +   
{\frac{u}{L^d}}
\sum_{\bf k_1,k_2,k_3,k_4}
	\Delta_{\bf k_1,k_2,k_3,k_4} \phi_{{\bf k_1}}	\phi_{{\bf k_2}}	\phi_{{\bf k_3}}	\phi_{{\bf k_4}},
	\quad
\label{eq-actionexpandedFBC}
\end{eqnarray}
where the $\Delta_i$'s are momentum-conserving factors.
The difference between the quadratic terms of Eqs.(\ref{eq-actionexpanded}) and~(\ref{eq-actionexpandedFBC}) is the source for the difference in scaling between the pseudocritical shifts in the PBC and FBC cases~\cite{RuGa85}.
The quartic term {in Eq.(\ref{eq-actionexpandedFBC})} is dangerous only for the modes $\vac k {\in Q}$ which couple to $h$.
We henceforth refer to modes for which $u$ is dangerous 
({in particular}, the zero mode {at $T_c$ and $T_L$} for PBC's and modes with all odd $n_\alpha$ {at $T_L$} for FBC's)
as Q-modes and the remaining ones as Gaussian modes or G-modes.

We introduce the notation $m_{\vac k}$ to represent 
the contribution of a single mode $\vac k$ to the average magnetization.
Thus $m_{\vac k}=\langle\phi_{\vac k}\rangle=\langle \int \phi(\vac x)\psi_{\vac k}(\vac x) d^dx \rangle$, where $\psi_{\vac k}(\vac x)$ is the standing wave or the sine mode depending upon the BC's. 
The brackets indicate the thermal average with the Boltzmann weight corresponding to the action (\ref{eq-actionexpanded}) or (\ref{eq-actionexpandedFBC}).
The equilibrium magnetization is then $m=\langle \int d^dx \sum_{\vac k}\phi_{\vac k}\psi_{\vac k}(\vac x)  \rangle$. 
The Q-modes  acquire non-vanishing expectation values and have projections onto the equilibrium magnetization  $m$ \cite{RuGa85,WiYo14}.
The G-modes do not have such projections and were expected to exhibit standard FSS given by (\ref{FSSPictureStandard}) with Landau exponents (\ref{eqMFT2}).
This was supported by numerical evidence for the susceptibility Fourier modes 
$\chi_{\vac k\in{Q}}$ in \cite{WiYo14}.

{\emph{Either}} we study physical quantities which are related to the {Q-modes,} in which case the DIV has to be properly taken into account, {\emph{or}} we  analyze properties associated with {G-modes} for which {$u$ is not dangerous}.
In the latter case the exponents are those predicted by the RG at the Gaussian FP; these are  (\ref{eqRG1}), (\ref{eqRG2}), {\it and not Landau exponents}  (\ref{eqMFT1}), (\ref{eqMFT2}).
In the first case, on the other hand, the exponents are indeed MFT exponents  (\ref{eqMFT1}), (\ref{eqMFT2}), but the correlation length has the FSS behavior involving $\qq$  and Eq.(\ref{FSSPictureStandard}) has to be modified to {QFSS which, e.g.  for magnetization and susceptibility, reads} 
{as}
 \bey m(t,L^{-1})& \sim & m(t,0) [L^\qqq/\xi(t)]^{-\frac\beta\nu}\sim L^{-\frac{\qqq\beta}{\nu}}\sim L^{-\frac d4},\quad 
\label{FSSmkoppa}\\
 \chi(t,L^{-1})& \sim & \chi(t,0) [L^\qqq/\xi(t)]^{\frac\gamma\nu}\sim 
L^{\frac{\qqq\gamma}{\nu}}\ \sim \ L^{\frac d2}.
\label{FSSchikoppa}
 \eey
{The Q-modes $m_{\vac k\in{Q}}$ and $\chi_{\vac k\in{Q}}$ themselves also obey QFSS. For G-modes  this reduces to standard FSS:}
\bey 
 m_{\vac k {\in{G}}}(t,L^{-1}) & \sim  &  {m_{\vac k {\in G}}(t,0)} 
[L/\xi(t)]^{-(d-y_h)} \sim  L^{-\frac{d-\sigma}{2}}, 
\quad ~ \label{FSSmkoppanonzero}\\
 {\chi_{\vac k {\in G}}(t,L^{-1}) } &  \sim  & {\chi_{\vac k{\in} 
G}(t,0)} [L/\xi(t)]^{2y_h-d}\ \sim L^{\sigma}.
\label{FSSchikoppanonzero}
 \eey

Eq.(\ref{FSSchikoppanonzero}) agrees numerically with the result of Wittmann and Young  \cite{WiYo14} because the  combination of exponents  $2y_h-d$  equals  {{$\sigma$  and  coincides 
with $\gamma/\nu$ (for the SRIM, $\sigma = 2$).}}
However, there is no way to reconcile Eq.(\ref{FSSmkoppanonzero}) with the power 
{{$-\beta/\nu=-\sigma / 2$ ($=-1$ for the SRIM)}} for arbitrary $d$ above $d_c=2\sigma$.
The long-standing puzzle associated with FBC's is due to the fact that if the {{Q-mode quantities}} follow  Eqs.~(\ref{FSSmkoppa}) and (\ref{FSSchikoppa}) at $T_L$, they fail at $T_c$ and obey instead scaling given by Eqs.~(\ref{FSSmkoppanonzero}) and (\ref{FSSchikoppanonzero}) like  G-quantities.

\begin{table}[b]
\vspace{-5mm}
\caption{The partitioning of Fourier modes into dangerous and non-dangerous sectors of the model. 
FSS in the  {Q (DIV)} sector is given by Eqs.(\ref{FSSmkoppa}) and (\ref{FSSchikoppa}). 
For the {G (non-DIV)} sector, 
it  predicts (\ref{FSSmkoppanonzero})  and (\ref{FSSchikoppanonzero}) while Landau FSS  gives
$m \sim L^{-\beta / \nu} = L^{-\sigma / 2}$ and  $\chi \sim L^{\gamma / \nu} = L^{\sigma }$ from (\ref{FSSPictureStandard}). 
}
\begin{tabular}{|c|c|c|c|c|}
\hline
       & \multicolumn{2}{c|}{\textbf{PBC}} &  \multicolumn{2}{c|}{\textbf{FBC}} \\ \cline{2-5} 
       & \multicolumn{2}{c|}{$\displaystyle{{k = \frac{2\pi}{L}(n_1,\dots,n_d)}}$} &  \multicolumn{2}{c|}{$\displaystyle{{k = \frac{\pi}{L+1}(n_1,\dots,n_d)}}$} \\ \cline{2-5} 
       & {{$n_\alpha=0~\forall~\alpha$}}    & {{$n_\alpha \ne 0$ for any $\alpha$}}               & {{All $n_\alpha$ odd}}     & {{Any $n_\alpha$ even}}               \\ \hline
 $T_L$ & {Q-modes}   &   {G-modes} &  {Q-modes} & {G-modes}       \\ \hline 
 $T_c$ & {Q-modes}   & {G-modes}   & {G-modes} & {G-modes}     \\ 
\hline
\end{tabular}
\label{table1}
\end{table}

\begin{figure}[!t]
\includegraphics[width=0.8\columnwidth, angle=0]{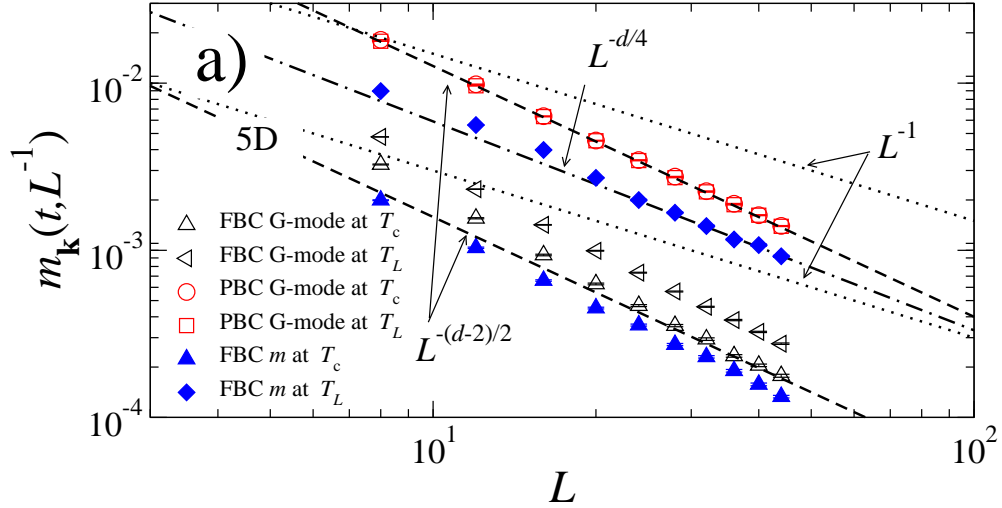}
\caption[a]{
{FSS for the 5D SRIM. 
The magnetisation for FBC's  belongs to the DIV sector and scales as $m(T_L) 
\sim L^{-d/4}$ ($\blue{\blacklozenge}$). 
At $T_c$ the  magnetization  scales as $L^{-(d-2)/2}$  as predicted by Gaussian 
FSS ($\blue{\blacktriangle}$).
The remaining data are for the lowest critical and pseudocritical G-modes 
$m_{\vac k}(L^{-1})$, each of which belongs to the non-DIV sector. 
Here we see they also follow Gaussian FSS and not the Landau FSS prediction 
$L^{-1}$ (which is indicated by the dotted lines). 
}
}
\label{fig:1}
\end{figure}

\begin{figure}[!t]
\includegraphics[width=0.8\columnwidth, angle=0]{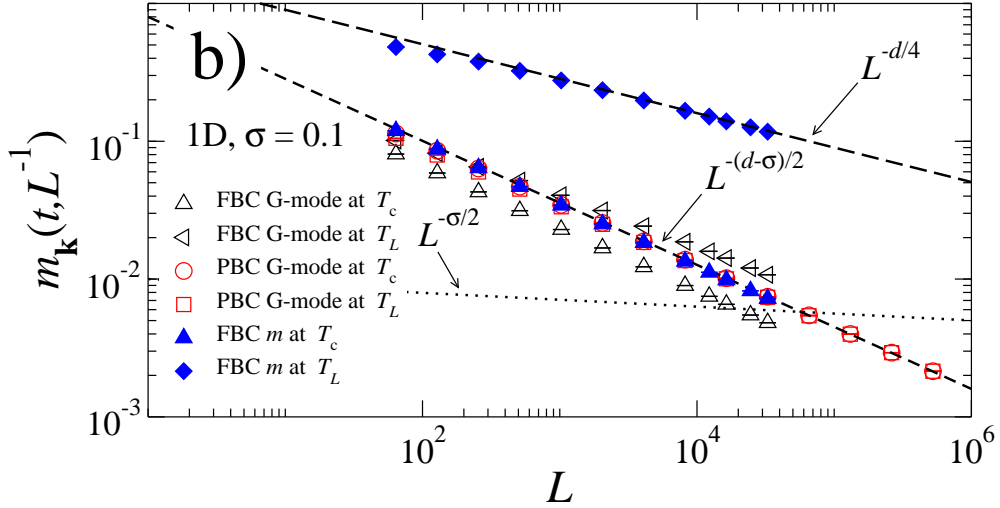}
\caption[a]{FSS for the  1D LRIM with $\sigma=0.1$. 
As in Fig.\ref{fig:1}, the magnetisation for FBC's exhibits QFSS at $T_L$,
scaling as $m(T_L) \sim L^{-d/4}$ ($\blue{\blacklozenge}$) and Gaussian FSS at 
$T_c$  scaling  as $L^{-(d-\sigma)/2}$  ($\blue{\blacktriangle}$).
The lowest critical and pseudocritical  G-modes all also follow Gaussian FSS
and not the Landau FSS prediction $L^{-\sigma / 2}$ (which is indicated by the 
dotted line).
}
\label{fig:2}
\end{figure}

In order to determine which of the forms 
(\ref{FSSmkoppa}) from QFSS, or (\ref{FSSmkoppanonzero}) from {the} Gaussian FP, 
or  
$m \sim L^{-\beta / \nu} = L^{-\sigma / 2}$ from Landau exponents in Eq.(\ref{FSSPictureStandard}),
correctly describes the scaling of the magnetization at $T_c$, we performed Monte Carlo simulations for the SRIM and the LRIM above $d_c$, along the lines described in \cite{ourEPJB}. 
The various sectors to be examined are summarised in Table~\ref{table1}.

Since we are simulating with finite-size lattices, we examine the magnitudes of the G-mode contributions through  {$m_{\vac k}(L^{-1}) 
=\langle{ | s_{\vac k} | }\rangle= \langle{| \sum_{\vac x} s_{\vac x} \psi_{\vac k}(\vac x) | }\rangle$, where $s_{\vac x}$ is an Ising spin at site $\vac{x=na}$.} 
The total magnetization is $m(L^{-1}) = \langle{|\sum_{\vac x}{s_{\vac x}}|}\rangle$.
For  the DIV sectors, scaling of the magnetization follows 
 $m(L^{-1})\sim L^{-d/4}$ {{from Eq.(\ref{FSSmkoppa}).
For PBC's at $T_c$, this is already well established, having also been verified numerically in \cite{Turks}.  We  find the same behaviour for PBC's at $T_L$.
Results for FBC's at $T_L$ 
support the same scaling there for both the SRIM and the LRIM. 
For the non-DIV sectors, standard FSS with Landau exponents predicts $m(L^{-1})\sim L^{-\sigma/2}$ {{while}} {we predict Gaussian scaling} $m(L^{-1})\sim L^{-(d-\sigma)/2}$ {{after Eq.(\ref{FSSmkoppanonzero}).}}
For the 5D SRIM, then, a plot on a double logarithmic scale should have slope {$-1$} according to ``Landau FSS'' or $-3/2$ according to {Eq.(\ref{FSSmkoppanonzero}). }
Our results  {in Fig.\ref{fig:1}}
are clearly compatible with the latter {{and rule out the former}}.
For {{the  LRIM above $d_c$}} the outcome is even more 
stark; ``Landau FSS'' predicts 
a  slope of $-\sigma/2 = -0.05$ {{independent of $d$}} 
and {Eq.(\ref{FSSmkoppanonzero})} predicts {$-(d-\sigma)/2=-0.45$, } 
for $d=1$,
the latter being 
clearly favoured by the data in {Fig.~\ref{fig:2}. }

 In the literature, it has been stated that ``due to the lack of a better way of treating the {{zero-momentum modes''}} it is usual to ``neglect them completely'' \cite{HeKa85}. 
Standard phenomenological FSS associated with the excited modes was then expected to deliver  ``Landau FSS''.
The results established here indicate that  this approach is not correct.
{Also in the literature, the critical exponents (\ref{eqRG1}) and (\ref{eqRG2}) are presented as non-physical; they were hitherto merely a step on the way to reconciling RG theory with Landau mean-field exponents above $d_c$. 
Here we have shown that they are, in fact, physically manifest in the magnitudes of the G-modes. This is not a feature of {finite size} only. In the thermodynamic limit, {G-}modes would, for example acquire the  temperature behaviour $m_{\vac k \in G}(t)\sim |t|^{d/2\sigma-1/2}$. 
Moreover, mean-field theory fails to fully describe either {G- or Q-modes} above the upper critical dimension. }
Instead  {this careful approach to RG, taking QFSS into account, delivers a consistent description for all Fourier modes.}

\vspace{0.5cm}
\begin{acknowledgements}
\noindent {\bf{Acknowledgements:}}
This work was supported by  EU FP7 IRSES projects 
and by the Coll\`ege Doctoral 
``Statistical Physics of Complex Systems'' Leipzig-Lorraine-Lviv-Coventry (${\mathbb L}^4$). 
\end{acknowledgements}

\end{document}